# Controlling and Streaking Nanotip Photoemission by Enhanced Single-cycle Terahertz Pulses

L. Wimmer*, G. Herink*, D. R. Solli, S. V. Yalunin, K. Echternkamp, and C. Ropers[1]

Institute of Materials Physics and Courant Research Center Nano-Spectroscopy and X-Ray Imaging, University of Göttingen, 37077 Göttingen, Germany, [1]Email: cropers@gwdg.de

**The active control of matter by strong electromagnetic fields is of growing importance, with applications all across the optical spectrum from the extreme-ultraviolet to the far-infrared. In recent years, phase-stable terahertz (THz) fields have shown tremendous potential in the observation and manipulation of elementary excitations in complex systems[1,2,3,4,5,6]. The combination of concepts from attosecond science with advanced THz technology facilitates novel spectroscopic schemes, such as THz streaking[7,8,9]. In general, driving charges at lower frequency enhances interaction energies[10,11,12] and can promote drastically different dynamics. For example, mid-infrared excitation induces field-driven sub-cycle electron dynamics in nanostructure near-fields[12,13]. Such frequency scalings will also impact nanostructure-based streaking, which has been theoretically proposed[14,15,16,17,18,19]. Here, we experimentally demonstrate extensive control over nanostructure photoelectron emission using single-cycle THz transients. The locally enhanced THz near-field at a nanotip significantly amplifies or suppresses the detected photocurrent. We present field-driven streaking spectroscopy with spectral compression and expansion arising from electron propagation within the nanolocalized volume. THz near-field streaking produces rich spectro-temporal features and will yield unprecedented control over ultrashort electron pulses for imaging and spectroscopy.**

Controlling electric charges with external fields is at the heart of modern information technology, with ultimate bandwidths limited by switching speeds in nanoscopic devices. The shortest electric field transients and most precise timings are reached in the optical domain[20,21,22]. The term *light-wave electronics* illustrates the anticipated application of optically field-driven processes to solids, starting from schemes initially developed for atoms and molecules[23]. In the THz range, strong table-top sources have opened up the field of nonlinear THz optics and are currently enabling comprehensive control over electronic and atomic motion, for example, in the manipulation of semiconductor charge carriers or spin waves, the triggering of phase transitions, or in molecular alignment and orientation[4,5,6,24,25,26].

*: These authors have contributed equally to this work.

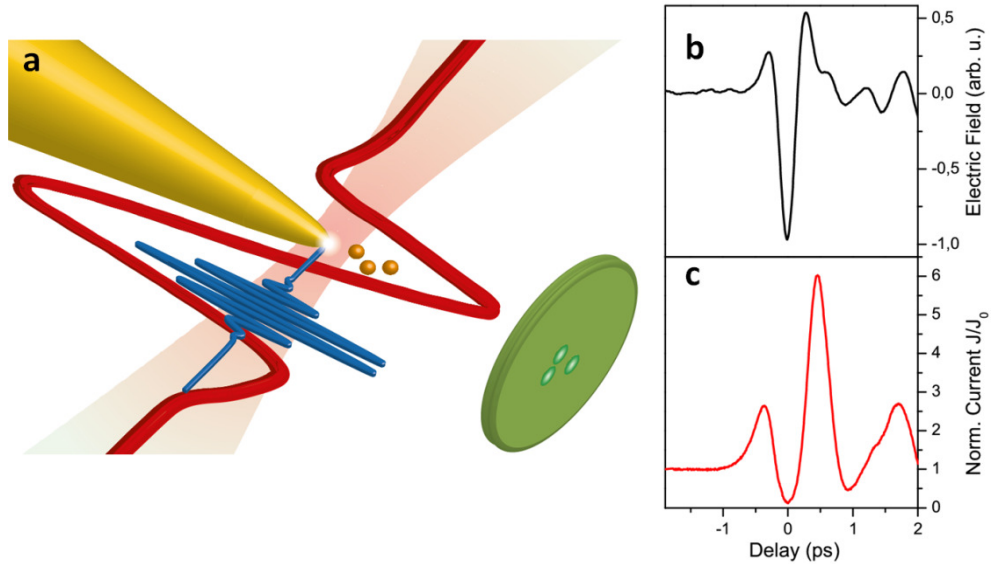

Figure 1: Gating nanotip photoemission with THz pulses. a: In the experiment, ultrashort THz and 800 nm pulses are focused onto a metallic nanotip. The photoemission current and spectrum are recorded as a function of relative pulse delay. b: Single-cycle THz transient measured by electro-optic sampling. c: Delay-dependent photocurrent, normalized to the photocurrent in the absence of the THz-pulse (tip bias: -10 V).

Completely new degrees of freedom are added by employing the enhancement and localization of optical fields within nanostructures. Specifically, at metallic nanotips, confined photoelectron emission[27,28] with characteristic strong-field features is observed[29,30], including carrier-envelope-phase (CEP) sensitivity of photoemission spectra[31]. In such structures, the spatial confinement of the near-field induces sub-cycle electron acceleration when driven at mid-infrared frequencies[12,13]. Strong and locally enhanced THz pulses, on the other hand, are expected to reveal the ultimate limits of field-driven sub-cycle dynamics in nanostructures. Yet, until now, the capabilities offered by intrinsically phase-stable THz generation and spectroscopy have not been applied to these systems. A variety of innovative space- and time-resolved streaking techniques were proposed in recent theoretical works[14,15,16,17,18,19]. A primary motivation for these schemes is to quantitatively map transient plasmonic fields by near-infrared streaking of UV-emitted photoelectrons. However, as we show in this Letter, nanostructure streaking concepts also offer means for far-reaching electron trajectory control, especially when translated to the THz spectral range.

Here, we demonstrate field-driven electron control based on dual-frequency excitation of a gold nanotip. Nonlinear photoelectron emission induced by near-infrared radiation at a nanotip apex is gated and streaked by locally enhanced single-cycle THz fields, and we find that the involved spatial, temporal and energetic scales are suitably matched for maximum tunability. The experiment is illustrated in Fig. 1**a** and represents a nanoscopic solid-state version of a THz streaking device. Nonlinear photoemission confined to the apex is driven by near-infrared (NIR) 50-femtosecond pulses at a center wavelength of 800 nm. These electrons are exposed to a THz-field of variable delay, which is generated in a light-induced air plasma[32,33]. The CEP-stable THz-transient (Fig. 1**b**) is characterized by electro-optic sampling in a ZnTe crystal (peak field of around 100 kV/cm). Both pulses are collinearly focused on a single electrochemically etched gold nanotip, and the electrons are detected with a microchannel-plate anode-assembly and a fast oscilloscope. Kinetic energy distributions are recorded using a retarding field analyzer.

The lightning-rod geometry of the sharp metal tip induces a local field enhancement, and as a result, moderate incident THz fields are capable of strongly modulating the near-infrared photoemission. In Fig 1**c**, the recorded photocurrent is shown as a function of the delay between the two pulses. We find that, depending on the phase of the THz field, the photocurrent is amplified by a factor of up to six or completely suppressed, thus allowing for effective switching that directly follows the THz-transient. We have observed such photocurrent switching for multiple tips, and the modulation depth sensitively depends on the tip curvature and associated field enhancement. The underlying mechanisms of this modulation involve both the THz-induced variation of the instantaneous surface electric field and with it the emission probability, as well as propagation effects of emitted electrons, which we will investigate below.

The temporally and spatially confined THz-field at the vicinity of the nanotip results in complex electron dynamics that are studied by recording energy spectra as a function of the relative pulse delay. This set of measurements forms a THz-streaking-spectrogram (Fig. 2). For negative delays, i.e., a situation in which the 800 nm pulse precedes the THz pulse, the photoemission spectrum is unaffected by the streaking field. The spectrum exhibits a peak shifted by the bias voltage, with an 8 eV width (FWHM) given by both strong-field induced broadening[34] and spectrometer resolution (limited to ~2 eV in this experiment).

In the delay range of temporal overlap between the NIR and THz pulses, the entire kinetic energy distribution is shifted in-phase with the photocurrent modulation (Fig. 2**b**). A comparison of the streaking spectrogram with the macroscopic electro-optic-sampling trace (Fig. 1**b**) demonstrates that overall, the momentary electric potential of the enhanced THz-field is directly translated into kinetic energy. We do not observe a purely linear spectral shift by the streaking field, as the tip geometry leads to spectrograms which are not symmetrically modulated for both THz-field polarities. Specifically, emitted electrons that are driven back to the tip may be reabsorbed or scattered at the surface. Thus, no electrons with energies far below the bias potential are found.

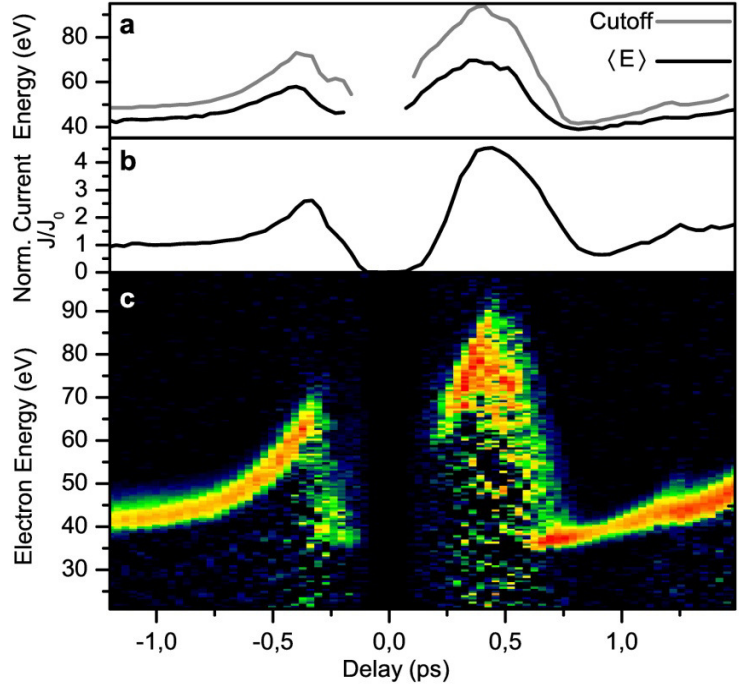

Figure 2: THz-Streaking spectroscopy of nanotip photoemission. a: Expectation value and cutoff energy of the delay-dependent kinetic energy spectra. b: Detected photoemission current, normalized to that in the absence of the THz pulse. c: Spectrogram composed of delay-dependent spectra (tip radius: 20 nm, bias: -40 V).

The fact that the spectrogram follows the THz field nearly in-phase is in stark contrast to previous implementations of streaking in diffraction-limited laser foci, including attosecond streaking[35] and THz-schemes to characterize X-ray and XUV pulses[7,9]. Under such conditions, the streaking field is spatially homogeneous, and photoemitted electrons, e.g., from an atomic gas, experience the entire evolution of the streaking field $F(t)$ after the time of emission $t_e$, resulting in the electron momentum governed by the temporal integral over the field, i.e., the vector potential: $p_{streak} = eA(t_e) = e\int_{t_e}^{\infty} F(t')\,dt'$ (35). In contrast, for the very sharp tip used in Fig. 2, the nanolocalized streaking geometry allows electrons to escape the enhanced near-field within a small fraction of the THz oscillation period, yielding energies $E_{streak}$ approximately given by the instantaneous electrical potential: $E_{streak} \approx eU(t_e) = e\int_{r=0}^{\infty} F(r, t_e)\,dr$. For a surface electric field decaying over a short length $l_F$ and in the quasi-static limit, the energy gained is directly proportional to $l_F eF(r = 0, t_e)$. With decay lengths of only few tens of nanometers and maximum energies of tens of electron volts, we thus observe strongly enhanced THz electric fields of multiple MV/cm at the tip. The associated sub-cycle reduction of the interaction time with local driving fields represents an extreme limit of the recently observed quenching of the quiver motion at nanostructures[12], and is phase-resolved here for the first time.

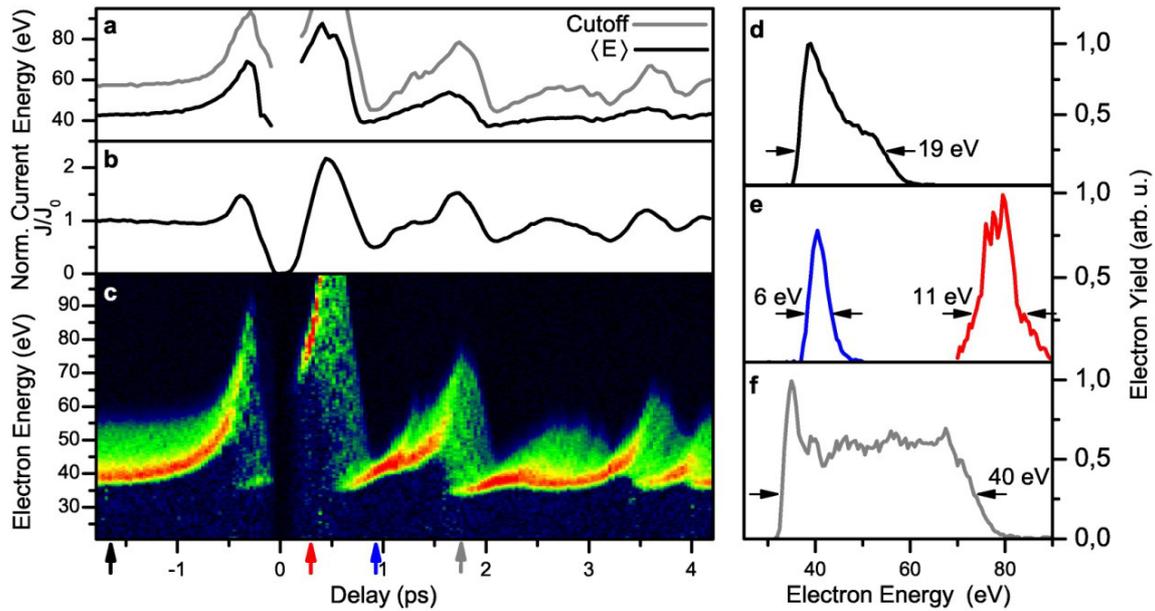

Figure 3: Spectral shaping observed in streaking spectrogroscopy. a: Expectation value and cutoff energy of the delay-dependent spectra. b: Normalized photoemission current. c: Spectrogram. d: Energy spectrum that is unaffected by the THz pulse (large negative delay). e: Compressed spectra. f: Expanded spectrum. Coloured arrows in c indicate delay times of respective spectra (tip radius: 40 nm, bias: -40 V).

As noted above, the spectrogram shows field-driven acceleration and quasi-instantaneous probing of the optical near-field. The spatial adiabaticity parameter $\delta = l_F m\omega^2/eF$ relates the near-field decay length to an electron's quiver amplitude in an oscillating field[12] and describes, how closely the final energy follows the instantaneous electric field ($\omega$: excitation frequency, $m$: electron mass). Generally, the present conditions imply quasi-static acceleration at the field maxima in the transient, with $\delta$-values far below unity (for example, $\delta = 0.02$ for $F$=2 MV/cm, $l_F = 20$ nm and at 1 THz frequency). However, especially closer to the zero-crossings of the field, longer interaction times and thus propagation effects become relevant. In particular, around the delay time, at which the photocurrent reappears after its central suppression ($\tau = 0.2$ ps), the spectral onset is immediately shifted to 60 eV, significantly above the energy supplied by the bias voltage. Thus, while the photocurrent quasi-instantaneously follows the local field, i.e., increases around zero field, the kinetic energy reflects an interaction over a small but noticeable integration time.

Such propagation effects arising from non-negligible interaction times can become rather pronounced as the tip radius is increased. In addition, they are revealed more distinctly by inducing a broader 'primary' NIR-photoemission spectrum, for example, realized by increased local intensity. Figure 3 displays a spectrogram where such conditions (larger tip radius; broader initial energy spectrum) are achieved. In the broader ensemble of initial energies generated by the 800 nm pulse alone (cf. Fig. 3**d**), electrons probe the spatially and temporally varying field over different time intervals. Specifically, faster electrons sample a more instantaneous THz-field as they leave the local hotspot at the apex, while slower electrons experience a time-varying transient. This has profound consequences in spectral redistributions, as evidenced by shape modulations in the spectrogram. As a general result, we find substantial THz-induced spectral broadening and compression (Figs. 3**e**,**f**) as inherent features of propagation within a localized and oscillating field. From a comparison of the current trace with the spectrogram, it is evident that compression and expansion are associated with rising and falling slopes, respectively, of the streaking force acting on the electrons (Figs. 4**e**,**f**). Compression in a temporally growing force follows from the fact that initially slower electrons, which interact over a longer period with the enhanced THz field, experience more overall acceleration than initially faster electrons, which leave the nanoscopic streaking volume more rapidly. Compression is apparent for various delays, e.g., at +1.0 ps and +2.0 ps. Spectral expansion, on the other hand, is a feature of a decreasing electric force: Fast electrons leave the high field quasi-instantaneously, acquiring maximum energy, while slower electrons experience both spatial and temporal field decay, potentially even being decelerated as the force changes sign, see, e.g., delays around -0.3 ps, +0.6 ps, +1.7 ps. We note that these findings represent a nano-optical equivalent of radio-frequency spectral and temporal compression of ultrafast electron pulses, yet at orders of magnitude smaller temporal and spatial scales[36,37].

Numerical modeling of the streaking spectrograms allows for a better understanding of the spatio-temporal acceleration processes and the identification of relevant physical parameters, such as the local THz field strength. Figure 4 presents results of a particle propagation simulation under conditions similar to those in the measurement of Fig. 3. In the model, electron trajectories exposed to the locally enhanced THz-field are computed for different emission times relative to the streaking-transient, and for a distribution of initial velocities (see methods). For simplicity, we separate the effects on the electron spectra induced by the NIR photoemission field and by the THz streaking field: The NIR-field is assumed to cause the broadening of the *initial* kinetic energy spectrum by strong-field emission and acceleration, as evident in the absence of the THz field at negative delays. In contrast, the THz-field provides the shift and reshaping of the *final* spectra after emission. This approach is justified by the more than two orders of magnitude differing frequencies at which emission and streaking occur. In addition, a modulation of the primary emission yield with the instantaneous THz-field for each delay is included in the model following separate measurements using a variable static field.

Figure 4**a** displays selected spectra simulated for various delay times, illustrating the strong spectral shaping induced by propagation in the spatially confined THz field (decay length 40 nm), see, for example, the spectral narrowing (blue, red) starting from the initially broader spectrum (black). The transfer function from initial to final energies for two selected emission times is plotted in Fig. 4**b**, while Fig. 4**c** shows the temporal evolution of a set of energies at two delay times resulting in spectral broadening (green) or compression (red). The full delay-dependent spectrogram (Fig. 4**d**) reproduces all main features of the measurements, including the THz-field induced energy shift, current enhancement and suppression, as well as complex spectral shaping. Even subtle characteristics, such as the narrow low-energy lobe at small negative delays, are successfully reproduced. Some differences between the experiment and the simulation in terms of absolute delay time positions of particular features arise from an only coarse modeling of the actual THz transient, which also displays some post-oscillations from water vapor absorption. The simulations confirm that the primary sources of spectral reshaping are propagation effects in the enhanced and localized THz-field (cf. sketches in Figs. 4 **e**,**f**). The model computations allow us to quantitatively determine the magnitude of the local THz field. We obtain a peak field strength of 9 MV/cm, which amounts to a local field enhancement at the tip of about 90. At frequencies of only few THz, such high fields are very hard to achieve in the absence in local field enhancements[5], and are close to the threshold for THz-induced tunneling.

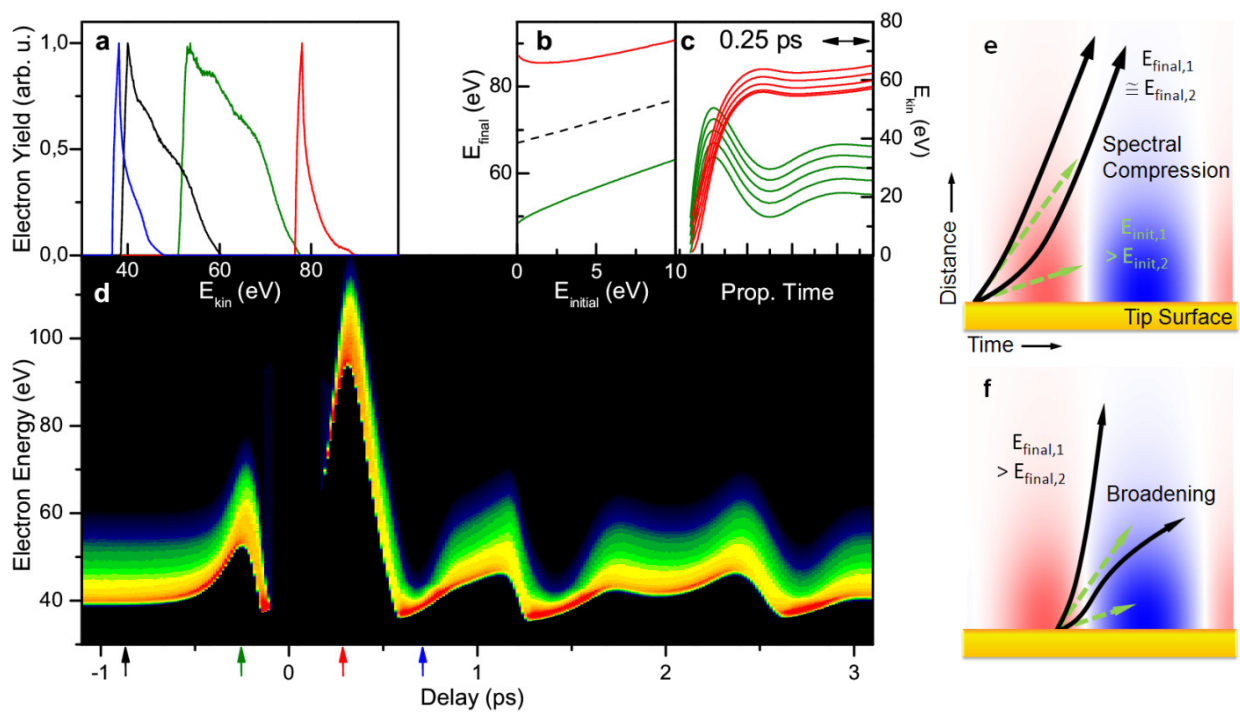

**Figure 4:** Simulation of THz-streaking at a nanotip. **a:** Energy spectra at different delay times. **b:** Final electron energy as a function of initial energy for two selected emission times corresponding to the spectra in **a**. Spectral compression and broadening are apparent (dashed line: unity slope for comparison). **c:** Temporal evolution of a set of five initial energies for these two emission times. **d:** Simulated spectrogram. Color-coded arrows indicate selected delays used in **a-c** panels. **e, f:** Illustrations of spectral compression (**e**) and broadening (**f**) of two different initial energies (green) induced by temporally rising and falling near-fields, respectively (red/blue: positive/negative force).

In conclusion, we demonstrate high-contrast switching and control of photocurrents and photoelectron spectra by tip-enhanced single-cycle THz-pulses. A moderate further increase in field strength will induce THz tunneling with new means to access excited state dynamics at surfaces and nanostructures. The spectral shaping and compression shown here amounts to a flexible tuning of the electron phase space density, which has great potential for use in ultrafast electron microscopy and spectroscopy. This effective electron dispersion control will also enable temporal compression at variable distances from the tip for novel imaging schemes. The transition region between nearly instantaneous and temporally integrating interaction at moderately sharp tips or higher frequencies will allow for a quantitative mapping of the spatial and temporal THz field distribution. In essence, a sampling of the streaking force with electrons of varying initial energy carries sufficient information for a complete near-field reconstruction or tomography.

**Methods:**

**Spectrogram simulation**

Electron trajectories are simulated for varying initial velocity and emission time in the THz near-field, yielding final kinetic energies as a function of initial energy. For simplicity, we consider a one-dimensional propagation characterized by a single field decay length. A typical THz transient with a spectrum and carrier-envelope phase of the measured electro-optic sampling (EOS) trace was used. The spatial decay of the THz field and the static field along the tip axis is modeled using a $r^{-1}$-dependence, where $r$ is the distance from the tip, in agreement with numerical solutions of Maxwell's equations for the tip geometry:
$F(r) \propto \frac{F_{r=0} \cdot l_F}{l_F + r}$, with the field decay length $l_F = 40$ nm.
The bias-voltage induced static field is accounted for in the same way, using a maximum field $E_{stat}(r=0) = \frac{U_{bias}}{k_{geo} R}$ with the tip radius $R = 40$ nm and a geometrical factor $k_{geo} = 2.5$. A maximum THz field strength of 9 MV/cm has to be used to account for the experimentally observed spectral features in Fig. 3, e.g., the maximum kinetic energy.

The primary NIR-induced kinetic energy distributions are modeled after those measured in the absence of the THz field. The final energy distributions are then obtained by applying the computed energy transfer function $E_{final}(E_{init})$ to the initial distribution.

**Acknowledgement:**

We thank K. Reimann and M. Wörner for helpful discussions and S. Schnell, B. Schröder and A. Feist for technical support. We gratefully acknowledge funding by the Deutsche Forschungsgemeinschaft (Priority Program 1391 "Ultrafast Nanooptics" and ZuK 45/1).